\documentclass[twocolumn]{aastex63}
\usepackage{color}

\received{---}
\revised{---}
\accepted{---}

\submitjournal{ApJL}

\shorttitle{Red Transients from Newborn Black Holes}
\shortauthors{Tsuna et al.}

\begin{document}

\title{Intermediate Luminosity Red Transients by Black Holes Born from Erupting Massive Stars}

\correspondingauthor{Daichi Tsuna}
\email{tsuna@resceu.s.u-tokyo.ac.jp}

\author{Daichi Tsuna}
\affiliation{Research Center for the Early Universe (RESCEU), Graduate School of Science, The University of Tokyo, 7-3-1 Hongo, Bunkyo-ku, Tokyo 113-0033, Japan}
\affiliation{Department of Physics, Graduate School of Science, The University of Tokyo, 7-3-1 Hongo, Bunkyo-ku, Tokyo 113-0033, Japan}

\author{Ayako Ishii}
\affiliation{Research Center for the Early Universe (RESCEU), Graduate School of Science, The University of Tokyo, 7-3-1 Hongo, Bunkyo-ku, Tokyo 113-0033, Japan}
\affiliation{Max Planck Institute for Gravitational Physics (Albert Einstein Institute), Am M\"{u}hlenberg 1, Potsdam-Golm 14476, Germany}

\author{Naoto Kuriyama}
\affiliation{Research Center for the Early Universe (RESCEU), Graduate School of Science, The University of Tokyo, 7-3-1 Hongo, Bunkyo-ku, Tokyo 113-0033, Japan}
\affiliation{Department of Astronomy, School of Science, The University of Tokyo, 7-3-1 Hongo, Bunkyo-ku, Tokyo 113-0033, Japan}

\author{Kazumi Kashiyama}
\affiliation{Research Center for the Early Universe (RESCEU), Graduate School of Science, The University of Tokyo, 7-3-1 Hongo, Bunkyo-ku, Tokyo 113-0033, Japan}
\affiliation{Department of Physics, Graduate School of Science, The University of Tokyo, 7-3-1 Hongo, Bunkyo-ku, Tokyo 113-0033, Japan}

\author{Toshikazu Shigeyama}
\affiliation{Research Center for the Early Universe (RESCEU), Graduate School of Science, The University of Tokyo, 7-3-1 Hongo, Bunkyo-ku, Tokyo 113-0033, Japan}
\affiliation{Department of Astronomy, School of Science, The University of Tokyo, 7-3-1 Hongo, Bunkyo-ku, Tokyo 113-0033, Japan}

\begin{abstract}
We consider black hole formation in failed supernovae when a dense circumstellar medium (CSM) is present around the massive star progenitor.
By utilizing radiation hydrodynamical simulations, we calculate the mass ejection of blue supergiants and Wolf-Rayet stars in the collapsing phase and the radiative shock occurring between the ejecta and the ambient CSM.
We find that the resultant emission is redder and dimmer than normal supernovae (bolometric luminosity of $10^{40}$--$10^{41}\ {\rm erg\ s^{-1}}$, effective temperature of $\sim 5\times 10^3$ K, and timescale of $10$--$100$ days) and shows a characteristic power-law decay, which may comprise a fraction of intermediate luminosity red transients (ILRTs) including AT 2017be. 
In addition to searching for the progenitor star in the archival data, we encourage X-ray follow-up observations of such ILRTs $\sim 1\mbox{-}10\,{\rm yr}$ after the collapse,  targeting the fallback accretion disk.
\end{abstract}


\section{Introduction} 
The main channel by which black holes (BHs) form is considered to be the gravitational collapse of massive stars. Numerical studies agree that a progenitor with a compact inner core can fail to revive the bounce shock (e.g. \citealt{OConnor11}).
These failed supernovae will be observed as massive stars suddenly vanishing upon BH formation, but their observational signatures are not completely known. Previous studies imply that the outcome depends on the star's angular momentum. Failed supernovae of stars with moderate or rapid rotation are expected to form accretion disks around the BHs, from which energetic transients are generated (\citealt{Bodenheimer83, Woosley93, MacFadyen99, Kashiyama15, Kashiyama18}; see also \citealt{Quataert19}).

For the dominant slowly rotating case, the collapsing star can still leave behind a weak transient after BH formation. During the protoneutron star phase before BH formation, neutrinos carry away a significant fraction of the energy of the core, around $10\%$ of its rest mass energy (e.g., \citealt{OConnor13}). This results in a decrease in the core's gravitational mass, and generates a sound pulse, which can eventually steepen into a shock and unbind the outer envelope of the star upon shock breakout \citep{Nadyozhin80,Lovegrove13,Fernandez18,Coughlin18a,Coughlin18b}.

\begin{figure*}
\centering
\includegraphics[width=0.85\linewidth]{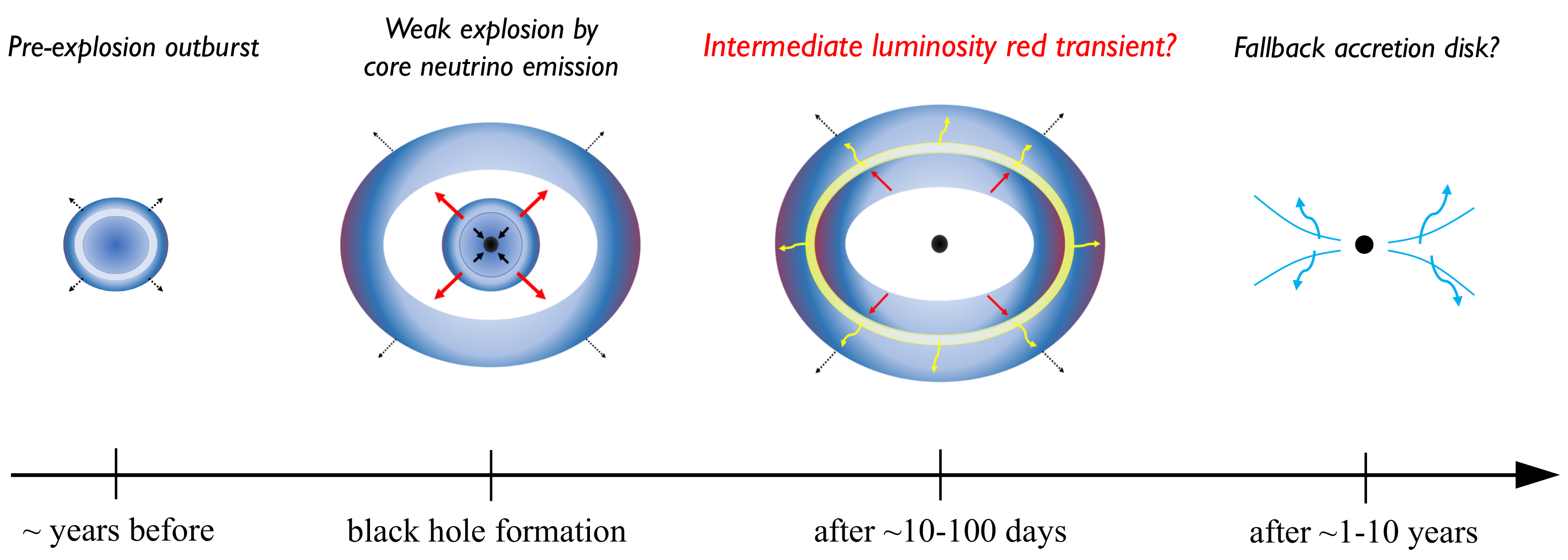}
\caption{Schematic figure of the emission we consider in this work. The ejecta created upon BH formation collides with the CSM created from a mass eruption $\sim$ years before core collapse. The kinetic energy of the ejecta is efficiently converted to radiation, being observable as intermediate luminosity red transients. The fallback of the outer layers of the envelope may form an accretion disk, that can be observable by X-rays $1$--$10$ yr after core collapse.}
\label{fig:ponchie}
\end{figure*}

This work investigates the emission from mass ejection of blue supergiant (BSG) and Wolf-Rayet (WR) progenitors that fail to explode. As shown in Figure \ref{fig:ponchie}, we particularly consider emission from the interaction between the ejecta and a dense circumstellar medium (CSM), which is commonly introduced to explain Type IIn supernovae (e.g. \citealt{Grasberg86}).
In our case, where the ejecta mass is expected to be much lighter than normal supernovae, CSM interaction is still, and even more, important to efficiently convert the kinetic energy into radiation
\footnote{For red supergiants, hydrogen recombination in the ejecta is expected to dominantly power the emission \citep{Lovegrove13}, and a strong candidate is already found \citep{Adams17a}.}.

We find that the resulting emission is similar to what has been classified as intermediate luminosity red transients (ILRTs) observed in the previous decades \citep{Kulkarni07,Botticella09,Smith09,Berger09,Bond09,Cai18,Jencson19,Williams20,Stritzinger20}. The origin of ILRTs is unknown, with several interpretations such as electron-capture supernovae or luminous blue variable-like mass eruptions. We propose an intriguing possibility that BH formation of massive stars can explain at least a fraction, if not all, of ILRTs.

This Letter is constructed as follows.
In Section 2 we present the details of our emission model and demonstrate that an ILRT AT 2017be can be naturally explained with our model. In Section 3 we estimate the detectability of these signals by present and future optical surveys, and suggest ways to distinguish this from other transients.

\section{Our Emission Model}
\subsection{The Dense CSM}
Past observations of Type IIn SNe found that the massive star's final years can be dramatic, with mass-loss rates of $10^{-4}$ -- $1\ M_\odot\ {\rm yr}^{-1}$ \citep{Kiewe12,Taddia13}. Such mass loss may be common also for Type IIP supernovae ($\gtrsim 70\%$; \citealt{Morozova18}), which comprises about half of all core-collapse supernovae. Although the detailed mechanism is unknown, the huge mass loss implies energy injection greatly exceeding the Eddington rate, or instantaneous injection with timescales shorter than the outer envelope's dynamical timescale.

\begin{figure*}
\centering
\begin{tabular}{cc}
\begin{minipage}{0.5\hsize}
\centering
\includegraphics[width=\linewidth]{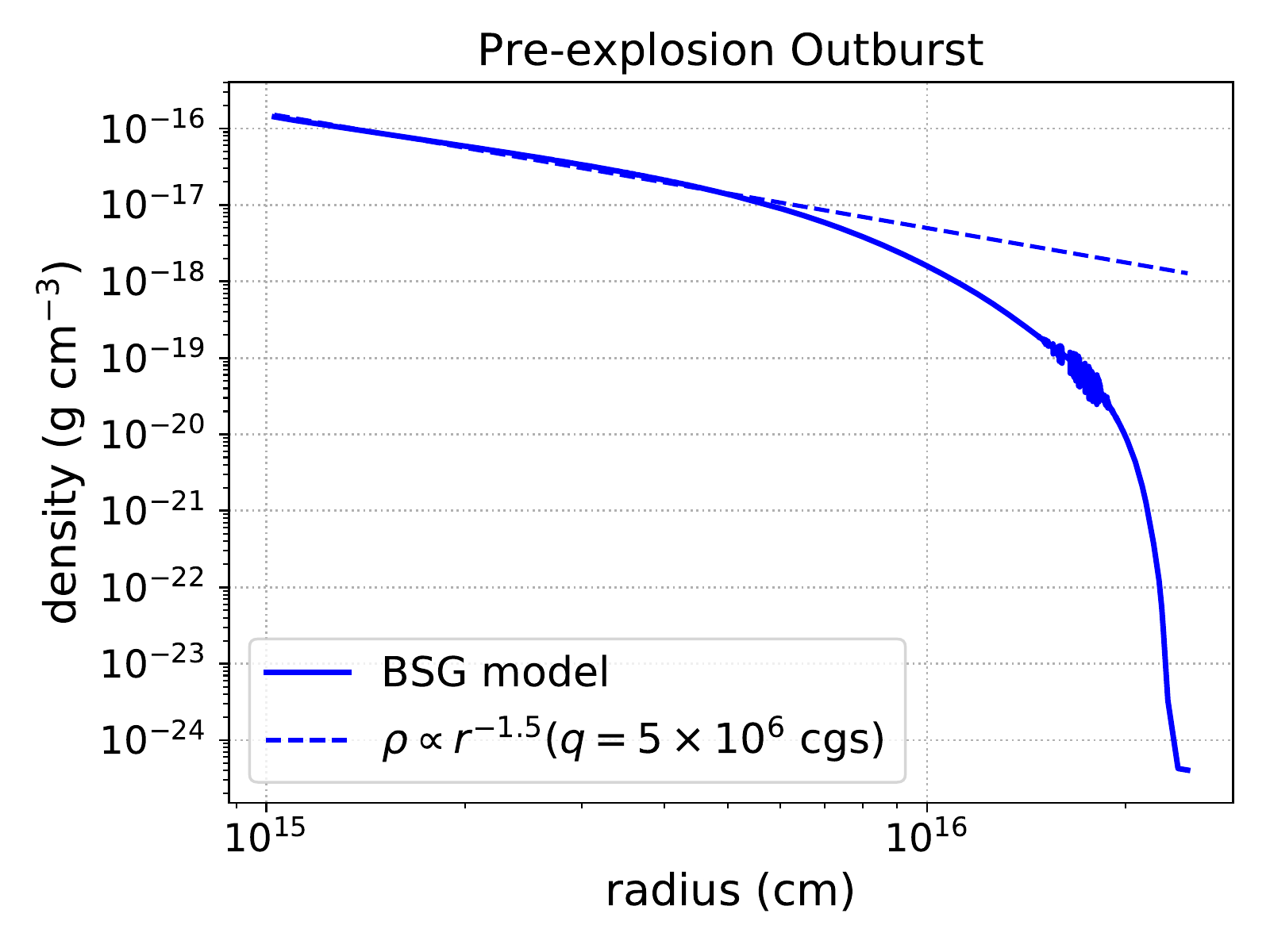}
\end{minipage}
\begin{minipage}{0.5\hsize}
\centering
\includegraphics[width=\linewidth]{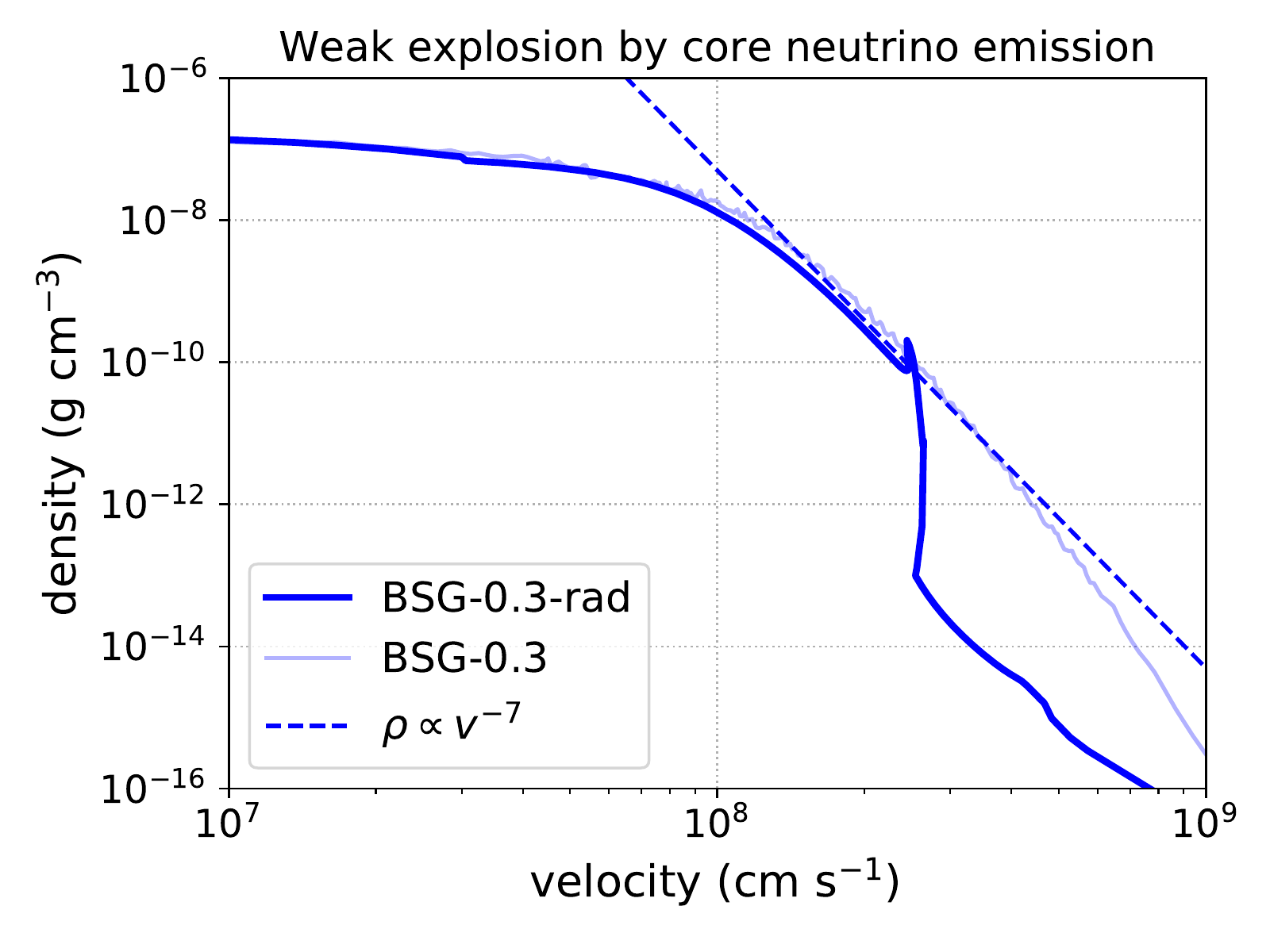}
\end{minipage}
\end{tabular}
\caption{(Left panel): Example density profile at core collapse of the unbound CSM that was erupted by energy injection 8.6 yr before core collapse. For this calculation, energy of $3.0 \times 10^{48}$ erg is injected in the base of the envelope for a timescale of $1.9 \times 10^4$ seconds, and the resulting total mass ejected is $1.9\times 10^{-2}\ M_\odot$. (Right panel): Ejecta density profile for the ``BSG--0.3" (light blue line, from the hydrodynamic simulation) and ``BSG--0.3--rad" (dark blue line, from the radiation hydrodynamic simulation). We use the data at $10^5$ s after core collapse, when the ejecta have become nearly homologous.}
\label{fig:BSG_CSM_ejecta}
\end{figure*}

\citet{Kuriyama19} studied the CSM resulting from energy injection, for various progenitors while being agnostic of the injection model. A notable finding was that the inner part has a profile of roughly $\rho(r)\propto r^{-1.5}$, shallower than the commonly adopted wind profile ($\rho(r)\propto r^{-2}$). An example is in Figure \ref{fig:BSG_CSM_ejecta}, where we plot the density profile of a BSG erupting a mass of $0.02\ M_\odot$ at $8.6$ yr before core collapse. Motivated by this, we assume the CSM profile to be a power law $\rho(r)=qr^{-s}$, and take $s=1.5$ as a representative value. The profile may be even shallower inwards due to the stronger gravitational pull from the central star. However, our assumption should not severely affect the dynamics of CSM interaction (given that the ejecta is heavier than the inner CSM), since the CSM mass is concentrated at where $s=1.5$. 

As neither the velocity nor mass-loss rate is constant, the standard parameterization of using wind velocity and mass-loss rate is not appropriate. Instead, we parameterize the dense CSM with its mass $M_{\rm CSM}$, and time $t_{\rm CSM}$ between its eruption and core collapse. Since the outer edge of the CSM is $v_{\rm out}t_{\rm CSM}$, where $v_{\rm out}$ is the velocity of the outer edge, the CSM mass is related to $q$ by $q\approx M_{\rm CSM}(3-s)(v_{\rm out}t_{\rm CSM})^{s-3}/4\pi$. Because the CSM is created from the marginally bound part, $v_{\rm out}$ is of order the escape speed at the stellar surface. For $s=1.5$,
\begin{equation}
q\sim 10^7\ {\rm cgs}\left(\frac{M_{\rm CSM}}{10^{-2}M_\odot}\right) \left(\frac{v_{\rm out}t_{\rm CSM}}{5\times 10^{15}\ {\rm cm}}\right)^{-1.5}.
\label{eq:q_vs_M_t}
\end{equation}
\citet{Kuriyama19} finds for their BSG and WR (their WR-1) models $q\sim 10^6$--$10^9$ cgs and $v_{\rm out}t_{\rm CSM}\sim 5\times 10^{15}$ cm.

The optical depth is given as a function of radius by $\tau(r) \approx \kappa q r^{1-s}/ (s-1)$, where $\kappa$ is the opacity. The photospheric radius where $\tau=1$ is at $r_{\rm ph, CSM} \approx \left[\kappa q/(s-1)\right]^{1/(s-1)}$. For $s=1.5$,
\begin{equation}
r_{\rm ph, CSM}\sim 2\times 10^{13}\ {\rm cm} \left(\frac{\kappa}{0.2\ {\rm cm^2g^{-1}}}\right)^2\left(\frac{q}{10^7\ {\rm cgs}}\right)^2.
\end{equation}
For a low value of $q$ that $r_{\rm ph, CSM}$ is smaller than the progenitor's radius, $r_{\rm ph, CSM}$ would instead be at the progenitor’s surface.

\subsection{Mass Ejection upon BH Formation}
To model the mass ejection at the BH formation, we conduct hydrodynamical and radiation hydrodynamical calculations on the response of the pre-supernova star to core neutrino mass loss. The details of the simulations are in the Appendix. We summarize the adopted parameters and resulting ejecta in Table \ref{tab:model}. 

\begin{table*}
\begin{tabular}{c|cccccccc}
\hline\hline
Name & $R_{\rm cc}$ & $M_{\rm cc}$ & $R_{\rm in}$ & $M_{\rm in}$ & $\delta M_G$ & $M_{\rm ej}$ [$M_\odot$]  & $E_{\rm ej}$ [erg] & $n$ \\
\hline
BSG--0.2 & $6.7\times 10^{12}$ & $11.7$ & $1.7 \times 10^9$& $3.9$ & $0.2$ & $0.076$  & $4.0\times 10^{47}$&10\\
BSG--0.3 & & &  & & $0.3$ & $0.11$  & $1.1\times 10^{48}$&10\\
BSG--0.4 & & &  & & $0.4$ & $0.16$ & $2.2\times 10^{48}$&10\\
BSG--0.3--rad & & & & & $0.3$ & $0.096$  & $6.0\times 10^{47}$&7\\
WR--0.3 & $2.9\times 10^{10}$ & $10.3$ & $2.2\times 10^9$ & $8.9$ & $0.3$ & $4.0\times 10^{-4}$   & $1.9\times 10^{46}$&10\\ 
\end{tabular}
\caption{Summary of the pre-supernova models and the resulting ejecta properties obtained from the simulations. The radius and mass at core collapse are $R_{\rm cc}$ and $M_{\rm cc}$ respectively. The inner boundary radius $R_{\rm in}$ and the enclosed mass $M_{r, \rm in}$ are set at where the gravitational timescale is approximately equal to the neutrino emission timescale. $\delta M_G$ is the gravitational mass loss by neutrinos. $M_{\rm ej}$, $E_{\rm ej}$, $n$ are the ejecta mass, energy, and power-law index of the density profile of the outer ejecta respectively.}\label{tab:model}
\end{table*}

We find that the ejecta properties are roughly consistent with \citet{Fernandez18}, with a double power-law density profile
\begin{eqnarray}
\rho (r,t)
&=& \left\{ \begin{array}{ll}
t^{-3}\left[r/(gt)\right]^{-n} & (r/t > v_t, \ {\rm outer\ ejecta}),\\
t^{-3}(v_t/g)^{-n} \left[r/(tv_t)\right]^{-\delta}  & (r/t < v_t,\ {\rm inner\ ejecta}),
\end{array}\right.
\label{eq:rho_ej}
\end{eqnarray}
where $g$ and $v_t$ are given by the ejecta mass $M_{\rm ej}$ and energy $E_{\rm ej}$ as

\begin{eqnarray}\label{eq:coeff_ej}
g &=& \left\{\frac{1}{4\pi(n-\delta)} \frac{[2(5-\delta)(n-5)E_{\rm ej}]^{(n-3)/2}}{[(3-\delta)(n-3)M_{\rm ej}]^{(n-5)/2}}\right\}^{1/n},\\
v_t &=& \left[\frac{2(5-\delta)(n-5)E_{\rm ej}}{(3-\delta)(n-3)M_{\rm ej}}\right]^{1/2}.\label{eq:v_t}
\end{eqnarray}
A notable point is that while overall we find from the hydrodynamical simulations that $n\approx10$ (consistent with that of successful supernovae; \citealt{MM99}), we find from radiation hydrodynamical simulations (``BSG--0.3--rad" model) that $n\approx 7$, and that the truncation of the outer ejecta for $v>2500\ {\rm km\ s^{-1}}$. This is due to the leaking of radiation incorporated in the ``BSG--0.3--rad" model, which prevents the radiative shock from pushing the ejecta to highest velocities upon shock breakout. Due to the truncation at the transition region between the inner and outer ejecta, a shallower profile of the outer ejecta is obtained. We thus adopt $n=7$ as a representative value, although stronger shocks may realize a larger $n$ close to the adiabatic case $n\approx10$.

\subsection{Emission from Ejecta-CSM Interaction}
Collision of ejecta and CSM creates forward and reverse shocks that heat the ambient matter and generate photons via free-free emission. When it is the outer ejecta component (of $r/t>v_t$) that pushes the shocked region, the shock dynamics can be obtained from self-similar solutions \citep{Chevalier82}. This solution considers collision between a homologous ejecta of profile $\rho(r,t)=t^{-3}(r/gt)^{-n}$, and CSM of profile $\rho(r)=qr^{-s}$ whose velocity is assumed to be negligible compared to the ejecta. The radius and the velocity of the contact discontinuity are obtained as
\begin{eqnarray}
r_{\rm sh}(t) &=& \left(\alpha g^n/q\right)^{1/(n-s)}t^{(n-3)/(n-s)}, \\
v_{\rm sh}(t) &=& (n-3)/(n-s)\cdot r_{\rm sh}/t,
\end{eqnarray}
where $\alpha$ is a constant determined by $n$, $s$, and the adiabatic index $\gamma$ assumed to be constant of $r$ and $t$.

At each shock's rest frame, kinetic energy of matter crossing the shock becomes dissipated, with some fraction $\epsilon$ converted to radiation. 
As the two shocks propagate outwards, free-free emissivity by shock-heated matter is reduced. Thus at early phases when enough photons can be supplied $\epsilon\sim 1$, while at later phases $\epsilon$ should drop with time. The boundary is different for the two shocks, as the densities in the two shocks' downstreams are usually much different \citep{Chevalier82}. 

An analytical model of the bolometric light curve incorporating this time-evolving efficiency but neglecting photon diffusion was developed in \citet{Tsuna19}. The bolometric light curve is a sum of two broken power laws (corresponding to two shocks), with indices given by $n$ and $s$ as
\begin{eqnarray}
L_{\rm bol}\propto 
\left\{ \begin{array}{ll}
t^{(-ns+2n+6s-15)/(n-s)} & (\epsilon \sim 1),\\
t^{(-2ns+3n+8s-15)/(n-s)} & (\epsilon<1),
\end{array}\right.
\end{eqnarray}
where the difference comes from the time dependence $\epsilon\propto t^{(-ns+n+2s)/(n-s)}$ when $\epsilon<1$. We refer to \citet{Tsuna19} for the full derivation of this luminosity.

\begin{figure*}
\centering
\begin{tabular}{cc}
\begin{minipage}{0.5\hsize}
\centering
\includegraphics[width=\linewidth]{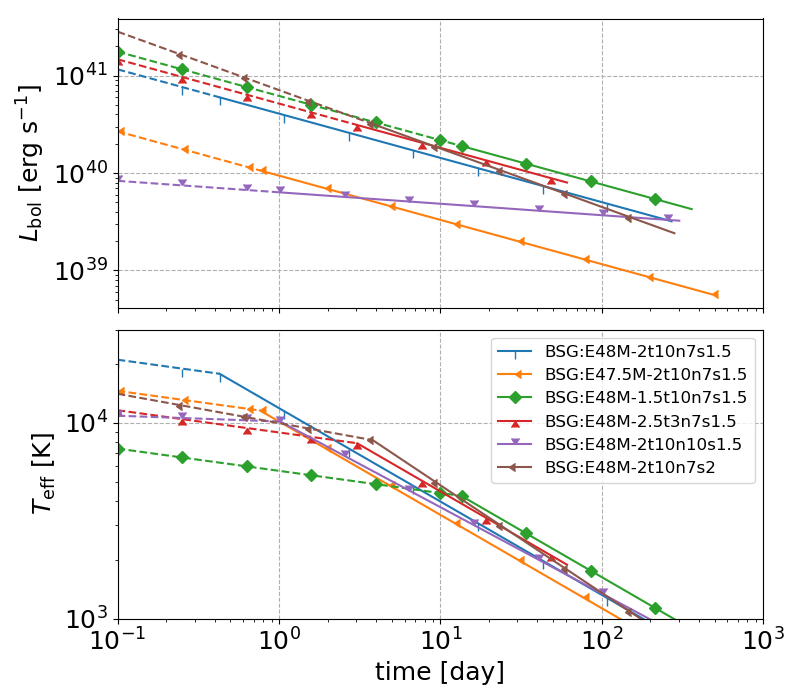}
\end{minipage}
\begin{minipage}{0.5\hsize}
\centering
\includegraphics[width=\linewidth]{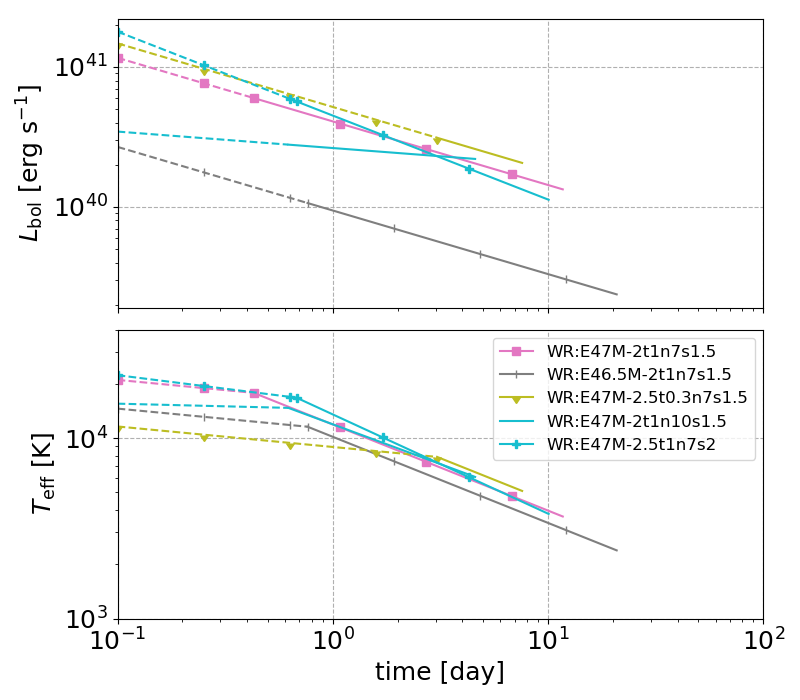}
\end{minipage}
\end{tabular}
\caption{Light curves for BSG and WR progenitors using our semianalytical model. The power-law feature extends until when either the reverse shock reaches the inner ejecta (i.e. $v_{\rm sh}=v_t$), or the shock reaches radius $v_{\rm out}t_{\rm CSM}$. The range shown as dashed lines is where diffusion can be important, which modifies the light curve from a power law.}
\label{fig:semiana}
\end{figure*}

Using this model we obtain bolometric light curves for various parameter sets of the ejecta and CSM, as shown in the top panels of Fig \ref{fig:semiana}. We follow \citet{Tsuna19} and set $\delta=1$ and $\gamma=1.2$. We define ``BSG-E$x$M$y$t$z$n$v$s$w$" as a BSG model with $E_{\rm ej}=10^{x}$ erg, $M_{\rm CSM}=10^{y}\ {\rm M_\odot}$, $t_{\rm CSM}=z$ yr, $n=v$, $s=w$, and similarly for WR models. The value of $q$ is obtained from equation \ref{eq:q_vs_M_t}, where we adopt $v_{\rm out}=10^2\ {\rm km\ s^{-1}}$ for BSG progenitors and $v_{\rm out}=10^3\ {\rm km\ s^{-1}}$ for WR progenitors. We also fix $M_{\rm ej}$ and adopt $0.1\ M_\odot$ for BSGs and $10^{-3}\ M_\odot$ for WRs. We obtain the mean mass per particle $\mu$ using the pre-supernova surface abundance listed in \cite{Fernandez18}, and set $\mu=0.846$ for BSGs and $\mu=1.84$ for WRs.

For early times when $r_{\rm sh}<r_{\rm ph, CSM}$, photons from the shocked region diffuse through the CSM with timescale
\begin{eqnarray}
t_{\rm diff} \sim \int_{R_{\rm cc}}^{r_{\rm ph, CSM}} dr\ \frac{\tau(r)}{c}.
\end{eqnarray}
where $R_{\rm cc}$ is the progenitor radius at core collapse. The shock reaches $r_{\rm ph, CSM}$ at time 
\begin{equation}
t_{\rm ph} \sim \left(\frac{\alpha g^n}{q}\right)^{-1/(n-3)}\left(\frac{\kappa q}{s-1}\right)^{(n-s)/[(s-1)(n-3)]}.
\end{equation}
The analytical model is valid only for $t \gtrsim {\rm min}(t_{\rm diff},t_{\rm ph})$ as it neglects diffusion. The regime where $t < {\rm min}(t_{\rm diff},t_{\rm ph})$ is plotted in Fig \ref{fig:semiana} as dashed lines. As hydrogen is reduced from solar abundance for these progenitors, we assume $\kappa=0.2\ {\rm cm^{2}\ g^{-1}}$. The border is found to be $<$ day for our fiducial BSG (BSG:E48M-2t10n7s1.5) and WR (WR:E47M-2t1n7s1.5) cases. 

We crudely estimate the temperature of the emission. For $t<t_{\rm ph}$, the effective temperature is
$T_{\rm eff} \sim [L_{\rm bol}/(4\pi r_{\rm ph, CSM}^2 \sigma_{\rm SB})]^{1/4}$,
where $\sigma_{\rm SB}$ is the Stefan--Boltzmann constant. For $t>t_{\rm ph}$, assuming the shocked region is optically thick,
$T_{\rm eff} \sim [L_{\rm bol}/(4\pi r_{\rm sh}^2 \sigma_{\rm SB})]^{1/4}$.
We show the temperature evolution in the bottom panels of Fig \ref{fig:semiana}. We note that one cannot rely on the assumption $\kappa=0.2\ {\rm cm^{2}\ g^{-1}}$ at late phases when $T_{\rm eff}\ll 5000$ K, since hydrogen starts to recombine and make the shocked region optically thin. Afterwards the spectrum for the late phase should deviate from a thermal one, and instead depend on the emission spectrum from the shocked region.

For the fiducial BSG:E48M-2t10n7s1.5 case, the luminosity and temperature scale with energy and time as
\begin{eqnarray}
L_{\rm bol} &\sim& 1\times 10^{40}\ {\rm erg\ s^{-1}} \left(\frac{E_{\rm ej}}{10^{48}\ {\rm erg}}\right)^{14/11}\left(\frac{t}{10\ {\rm days}}\right)^{-5/11}\\
T_{\rm eff} &\sim& 4\times 10^3\ {\rm K}\left(\frac{E_{\rm ej}}{10^{48}\ {\rm erg}}\right)^{3/22} \left(\frac{t}{10\ {\rm days}}\right)^{-21/44}.
\end{eqnarray}
The dependence of $T_{\rm eff}$ on the CSM parameters is not simple, as the power-law index can change at around 10 days. Despite uncertainties in the precise temperature and spectrum, we find that the features (timescale of $10$--$100$ days, luminosity $\sim 10^{40}\ {\rm erg\ s^{-1}}$, temperature $\sim 5\times 10^3$ K at 10 days for our fiducial model) are similar to ILRTs.

The power-law feature of the CSM interaction and the resultant light curve is valid until either
    (i) the reverse shock reaches the inner ejecta or
    (ii) the forward shock reaches the outer edge of the dense CSM.
If either occurs, the dissipated kinetic energy and/or the radiation conversion efficiency drops, resulting in a cutoff in the light curve. Case (i) occurs when $v_{\rm sh}\approx v_t$, at 
\begin{equation}
t_{\rm core} \approx \left[\sqrt{\frac{2(5-\delta)(n-5)(n-s)^2E_{\rm ej}}{(3-\delta)(n-3)^3M_{\rm ej}}} \left(\frac{\alpha g^n}{q}\right)^{-1/(n-s)} \right]^{\frac{n-s}{s-3}}
\label{eq:t_core}
\end{equation}
For the fiducial BSG:E48M-2t10n7s1.5 model,
\begin{equation}
t_{\rm core} \sim 250\ {\rm days} \left(\frac{E_{\rm ej}}{10^{48}\ {\rm erg}}\right)^{-1/2}\left(\frac{M_{\rm CSM}}{10^{-2}{\rm M_\odot}}\right)^{-2/3} \left(\frac{t_{\rm CSM}}{10\ {\rm yr}}\right).
\end{equation}
Case (ii) occurs when $r_{\rm sh}=r_{\rm out}$, at
\begin{eqnarray}
t_{\rm out} \sim \left(\alpha g^n/q\right)^{-1/(n-3)}r_{\rm out}^{(n-s)/(n-3)}
\label{eq:t_out}
\end{eqnarray}
For the fiducial BSG:E48M-2t10n7s1.5 model,
\begin{equation}
t_{\rm out} \sim  270\ {\rm days}\ 
\left(\frac{E_{\rm ej}}{10^{48}\ {\rm erg}}\right)^{-1/2}
\left(\frac{M_{\rm CSM}}{10^{-2}{\rm M_\odot}}\right)^{1/4} \left(\frac{t_{\rm CSM}}{10\ {\rm yr}}\right).
\end{equation}
The two timescales may constrain $t_{\rm CSM}$, which may give us some information on the progenitor's activity just before core collapse.

\subsection{Comparison with Observed ILRTs}
Intriguingly, we may have already detected these events as ILRTs. We attempt to explain the observations of an ILRT AT 2017be \citep{Adams18} by our model with a BSG progenitor.

Previously \citet{Cai18} claimed that this transient is likely to be an electron-capture supernova (ECSN). However, compared to the luminosity of a few $\times 10^{40}\ {\rm erg\ s^{-1}}$ for AT 2017be, light curves from ECSN are predicted to have a plateau phase much brighter (of $\sim 10^{42}\ {\rm erg\ s^{-1}}$) lasting for 60-100 days (e.g. \citealt{Tominaga13,Moriya14b}). We propose that a failed supernova, which has a much weaker explosion than ECSN, can reproduce the observations of AT 2017be.

\begin{figure}
\centering
\includegraphics[width=\linewidth]{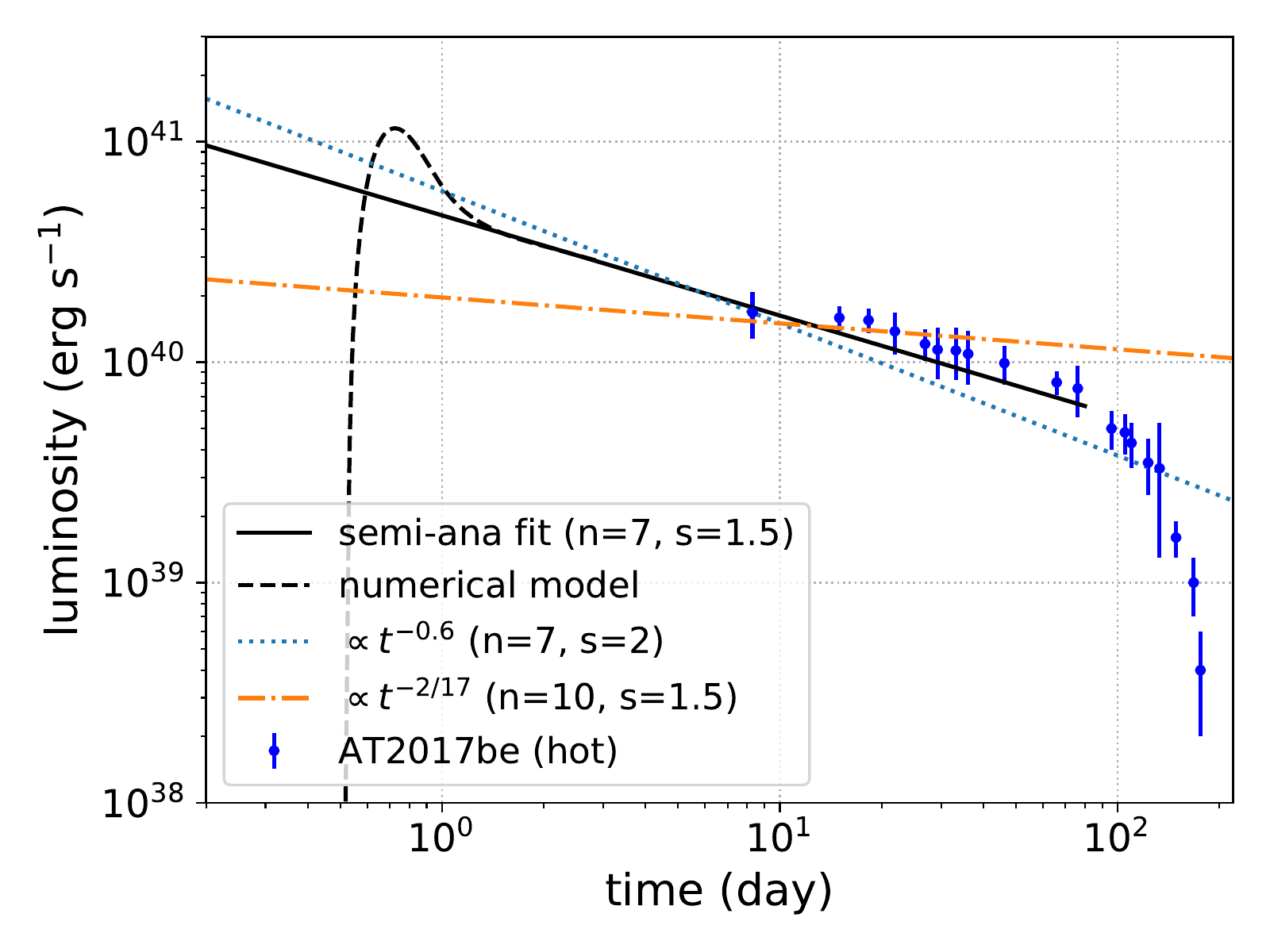}
\caption{Fit of our failed BSG model to the luminosity evolution of the hot component of AT 2017be. The solid (dashed) line is from the semianalytical (numerical) light curve model in \citet{Tsuna19}. The power-law indices $n=7$ and $s=1.5$ match AT 2017be better than other sets of $n$ and $s$ in the plot (assuming a time independent $\epsilon$) which predicts different power-law decays. Our model fit is stopped at $80$ days, when the light curve is expected to display a cutoff (see the main text).}
\label{fig:fitAT 2017be}
\end{figure}

We consider ejecta of mass $0.1\ M_\odot$, energy $5.3\times 10^{47}$ erg, and $n=7$ colliding with a CSM of $s=1.5$. We set $q=1.2\times 10^8$ cgs, corresponding to $M_{\rm CSM}$ of
\begin{eqnarray}
M_{\rm CSM} \sim 1.6\times 10^{-2}\ {\rm M_\odot} \left(\frac{v_{\rm out}t_{\rm CSM}}{10^{15}{\rm cm}}\right)^{3/2}
\end{eqnarray}
Figure \ref{fig:fitAT 2017be} shows the calculated light curve, with a comparison between that obtained in \citet{Cai18} from the "hot component" of their two-component SED fit\footnote{The explosion date is assumed to be 15 days before the observed r-band peak \citep{Cai18}.}. The solid black line is from the aforementioned analytical model, and the dashed black line is from a numerical model also introduced in \citet{Tsuna19} that takes into account diffusion. The numerical curve is just meant to be a demonstration of the validity of the analytical model. The exact luminosity and timescale at peak is questionable due to the likely absence of the fastest component in the ejecta (see Section 2.2), which is not taken into account in the simulation. Nonetheless the late phases of the light curve that we compare here should be robust.

We find that the power-law feature of the observed light curve is naturally reproduced by our model, which predicts profiles of $n=7$ and $s=1.5$, while other power-law curves based on different ejecta and CSM profiles do not. The best-fit values of $n$ and $s$ matching with what we expect {\it a priori} gives support to our model.

The observed light curve shows a cutoff from $\sim 100$ days. We test if this cutoff can be explained by our model. Plugging the values we assumed for the ejecta and CSM into equations \ref{eq:t_core} and \ref{eq:t_out}, we find
$t_{\rm core} \sim  80\ {\rm days}$
and
$t_{\rm out} \sim  130\ {\rm days} \left(v_{\rm out}t_{\rm CSM}/10^{15}{\rm cm}\right)^{11/8}$.
Thus we can explain the cutoff at around 100 days if $v_{\rm out}t_{\rm CSM}\gtrsim 10^{15}$ cm. This constraint and $v_{\rm out}$ of order $100\ {\rm km\ s^{-1}}$ implies that the mass eruption occurred years before core collapse. This timescale is consistent with proposed mechanisms for mass eruption \citep{Quataert12,Moriya14,Smith14}.

In fact, if this mass eruption was due to energy injection from the interior of the star, the eruption itself can be luminous, with peak luminosity of $10^{40}$--$10^{41}\ {\rm erg\ s^{-1}}$ \citep{Kuriyama19}. This may have been detectable, as was the case for outbursts observed years before the terminal explosion in, e.g. SN 2006jc \citep{Pastorello07} and 2009ip \citep{Mauerhan13}. Unfortunately pre-explosion images for AT 2017be are insufficient to test this or identify the progenitor star.

We compare our model to other ILRTs whose bolometric light curves were available, and find that AT 2019abn \citep{Williams20} can be consistent with our model. Its bolometric light curve shows a shallow power-law decay until $\sim 50$ days since discovery, followed by a nearly exponential cutoff. However, the power-law index ($\propto t^{-0.2}$--$t^{-0.3}$, considering the uncertainty on the explosion epoch) is shallower than that predicted from $n=7$ and $s=1.5$ ($\propto t^{-0.45}$). This can be reconciled by adopting a shallower $s$ of $0.3$--$0.9$, or steeper $n$ of $8$--$9$. The latter requires a stronger explosion to preserve the faster part of the ejecta. Qualitatively, this may also explain AT 2019abn having an order of magnitude higher luminosity than AT 2017be. Although some ILRTs have identified progenitors that disfavor a BSG or WR origin \citep{Prieto08,Thompson09,Kochanek11}, our model may comprise a nonnegligible fraction of ILRTs.

\section{Discussion}
We estimate the local event rate of these events. If we use the local (successful) core collapse supernova rate $R_{\rm SN} \sim 7\times 10^{-5}\ {\rm Mpc^{-3}\ yr^{-1}}$ \citep{Li11}, the all-sky event rate $R$ of failed supernovae within distance $d$ is
\begin{eqnarray}
R=f\times \frac{4\pi}{3} d^3 R_{\rm SN}\sim 0.8\ {\rm yr^{-1}} \left(\frac{f}{0.1}\right) \left(\frac{d}{30\ {\rm Mpc}}\right)^3,
\end{eqnarray}
where $f$ is the fraction of failed BSG/WR explosions among all core collapse. 

Overall, the signal has luminosity of order $10^{40}\ {\rm erg\ s^{-1}}$ and time scale of order 10 days. For a blackbody emission of 5000 K and luminosity $10^{40}\ {\rm erg\ s^{-1}}$, the AB magnitude in the V and R bands are $\approx -11$ mag. The Zwicky Transient Facility (ZTF; \citealt{Bellm19}) and the Legacy Survey of Space and Time (LSST; \citealt{Ivezic08}), with its survey having a sensitivity of $21$ and $25$ mag in these bands and cadence of 3 days, can in principle detect them out to $\sim 25$ Mpc and $\sim 100$ Mpc respectively.

Once a failed supernova candidate is found, it is important to distinguish this from other origins. A smoking gun may be the identification of the progenitor from archival data, as was done using archival Hubble Space Telescope images in \citet{Gerke15}. This may be possible if the source is out to $\lesssim 30$ Mpc \citep{Smartt09}.

Another smoking gun may be X-ray emission, if a fraction of the outer envelope falls back to the BH with sufficient angular momentum to create an accretion disk. \citet{Fernandez18} claims from simple estimates that this can occur for BSG and WR progenitors. Failed supernovae from these progenitors not only have large fallback but also small $M_{\rm ej}$, making them suitable targets to X-ray newborn BHs. Extrapolating the fallback rate obtained by \citet{Fernandez18} with the standard $\dot{M}\propto t^{-5/3}$ law, we find that the accretion rate persists above the Eddington rate of a $10M_\odot$ BH for $30$ years for BSGs and $1$ year for WRs.  For an X-ray emission of Eddington luminosity from a $10M_\odot$ BH, the flux is
$F_X \sim 1\times 10^{-14}\ {\rm erg\ s^{-1}} (d/30\ {\rm Mpc})^{-2}$, within reach for current X-ray telescopes once the ejecta become transparent to X-rays. The opacity to soft X-rays is mainly controlled by the photoelectric absorption by oxygen, whose cross section is $\sim 10^{-19}\ {\rm cm^2}$ at 1 keV for electrons in the K-shell. The {\it oxygen} column density of the ejecta is
\begin{eqnarray}
N_O &\sim& \frac{X_O M_{\rm ej}/16m_p}{4\pi (v_{\rm ej} t)^2} \nonumber \\
&\sim& 10^{21}\ {\rm cm^{-2}} \left(\frac{X_O}{0.01}\right)\left(\frac{M_{\rm ej}}{0.1M_\odot}\right)\left(\frac{v_{\rm ej}}{10^3{\rm km\ s^{-1}}}\right)^{-2}\left(\frac{t}{{\rm yr}}\right)^{-2}.
\end{eqnarray}
The oxygen mass fraction $X_O$ is $\sim 0.01$ for BSGs which roughly follow the solar abundance, but higher ($\sim 0.1$) for WR stars. For BSG (WR) ejecta of $M_{\rm ej}\sim 0.1M_\odot, v_{\rm ej}\sim 10^3\ {\rm km\ s^{-1}}$ ($M_{\rm ej}\sim 10^{-3}M_\odot, v_{\rm ej}\sim 3\times 10^3\ {\rm km\ s^{-1}}$), $N_O$ becomes $\sim 10^{19}\ {\rm cm^{-2}}$ at 10 (1) yr, and afterwards the ejecta are transparent to X-rays. We thus encourage X-ray follow-up observations of ILRTs detected in the past decades, including AT 2017be.

When the ejecta is still optically thick to X-rays, injection of X-ray (and possible outflow from the accretion disk) can heat the ejecta. This may rebrighten the ejecta, analogous to what is proposed in \cite{Kisaka16} in the context of neutron star mergers. The detailed emission will depend on when the fallback matter can create an accretion disk.)

\acknowledgments
We thank the anonymous referees for important comments that improved this manuscript. We thank Rodrigo Fern\'{a}ndez for providing us the pre--supernova stellar models, and also thank Kotaro Fujisawa and Yuki Takei for discussions. 
DT is supported by the Advanced Leading Graduate Course for Photon Science (ALPS) at the University of Tokyo. This work is also supported by JSPS KAKENHI Grant Numbers JP19J21578, JP17K14248, JP18H04573, 16H06341, 16K05287, 15H02082, 20K04010, MEXT, Japan.

\appendix
\section{Details of the Simulation of Mass Ejection upon BH Formation}
The hydrodynamical and radiation hydrodynamical calculations were done by Lagrangian codes developed in \citet{Ishii18} and \citet{Kuriyama19} respectively. For both of the two codes we use the same pre-collapse progenitor model as \citet{Fernandez18}, and have modeled the neutrino mass loss by reducing the gravitational mass in the core by $\delta M_G$, which is varied between the most conservative ($\sim 0.2\ M_\odot$) and most optimistic ($\sim 0.4\ M_\odot$) cases adopted in \citet{Fernandez18}. The outer boundary of the computational region is the progenitor's surface, and the inner boundary $r_{\rm in}$ is set to be the radius where the freefall timescale is equal to the timescale of neutrino emission $\tau_{\nu}$. This is justified by the fact that for matter at $r<r_{\rm in}$, the freefall timescale is so short that it fails to react to the gravitational loss by neutrino emission and is swallowed by the BH, whereas the matter at $r>r_{\rm in}$ has enough time to react. We determine $r_{\rm in}$ from the equation $\sqrt{r_{\rm in}^3/GM_{r,\rm in}} = \tau_{\nu}$, where $G$ is the gravitational constant and $M_{r,\rm in}$ is the enclosed mass within $r_{\rm in}$. Following \citet{Fernandez18}, we set $\tau_{\nu}$ to 3 s. We also follow \citet{Fernandez18} and include a prescription to remove the innermost cells that fall onto the core much faster than the local sound speed.

\bibliographystyle{apj} 
\bibliography{failedSN}

\end{document}